\documentclass[secnumarabic,amssymb,aps,pre,preprint]{revtex4-1}
\usepackage[dvipdfm]{graphicx}
\usepackage{amsmath}  
\usepackage{latexsym,mathrsfs}
\usepackage{dcolumn}
\usepackage{bm}

\usepackage{graphicx}
\usepackage{dcolumn}
\usepackage{bm}

\begin{document}

\title{Hierarchical logistic equation to describe the dynamical behavior of penetration rates}

\author{Tohru Tashiro}
\email{tashiro@cosmos.phys.ocha.ac.jp}   
\author{Hiroe Minagawa}
\author{Michiko Chiba}
\affiliation{Department of Physics, Ochanomizu University, 2-1-1 Ohtsuka, Bunkyo, Tokyo 112-8610, Japan}
\date{\today}

\begin{abstract}
We propose a hierarchical logistic equation as a model to describe the dynamical behavior of a penetration rate of a prevalent stuff. In this model, a memory, how many people who already possess it a person who does not process it yet met, is considered, which does not exist in the logistic model. As an application, we apply this model to iPod sales data, and find that this model can approximate the data much better than the logistic equation. 

\end{abstract}

\maketitle

\section{Introduction}

How does fashion diffuse in our society?
Fashion spreads out, although we, members of society,  do not aim to do so: Most of us are not out to spread it, while some may have that aim.
New stuffs that are somewhat of a curiosity at the beginning will become commonplace before we notice.  
Some of them may disappear from our life style.
This phenomenon is essentially similar to various changes of phase in matter which we cannot imagine from an interaction between atoms or molecules.
That is to say, the human being is ``the social atom''\cite{Buchanan07}.

We want to clarify the mechanism producing the occurrence of fashion. In order to deal with this as a scientific problem, quantitative data indicating the extent of diffusing are necessary. Here, we shall employ a {\em penetration rate}.  This has the following universality: Generally, penetration rates increase slowly at the beginning and, then, the growth reaches its maximum. Finally, the rates become saturated. This change in time is called the {S-shaped curve}. 
The logistic function, solution of the logistic equation, has often been used to analyze the rates. 
The logistic equation can be represented by the following differential equation.
\begin{equation}
\frac{{\rm d}x(t)}{{\rm d}t} = \frac{a}{X}x(t)\left\{X-x(t)\right\}
\label{eq:logi}
\end{equation}
This solution is the logistic equation:
\begin{equation}
x(t) = \frac{X}{1+\left\{\frac{X}{x(0)}-1\right\}{\rm e}^{-at}} \ .
\end{equation}
If $x(0) < X$, then, $x(t)$ is in an interval from $0$ to $X$ and forms the {S-shaped curve}.
Therefore, $x(t)/X$ can be employed for penetration rates since it does not extend beyond $1$.

As is well known, the logistic equation was proposed as an equation describing a population growth with an upper limit by Verhulst \cite{Verhulst38,Verhulst45,Verhulst47}.
However, the value of his study was not accepted in those days.
In 1920, about a century later, Pearl and Reed rediscovered this equation while investigating the evolution of fly population \cite{Pearl&Reed20}.
Lotka also derived this equation as the model of population growth \cite{Lotka25}.
Their works excavated the logistic equation. 

Griliches made the first adoption of the logistic equation for the dynamical behavior of innovation diffusion \cite{Griliches57}.
He analyzed the penetration rate of hybrid corn among farmers by the logistic function.
After that, Mansfiled justified using the logistic equation for the innovation diffusion mathematically \cite{Mansfield61}.
On the other hand, Fisher and Pry utilized this for a substitution of a share of two products,  i.e., margarine and butter, \cite{Fisher&Pry71}.

In this way, the employment of the logistic equation for the innovation diffusion started and, then, this has been utilized for dynamical behavior of penetration rates for various stuffs: in the past decade, mobile phones\cite{Gruber01,Frank04,Boretos07}, personal computers\cite{Liu06,Yang09,Dwivedy10,Yu10,Teng02}, electronics\cite{Liu06,Tasaki01,Yamasue06}, energy technologies\cite{Hafele81,Lund06}, information technologies\cite{Teng02} and oxygen-steel making process\cite{Kumar92}.

Thus, the change of many penetration rates in time can be described by the logistic function.
But, what kind of human communication results in such a dynamical behavior? It is not a self-evident question.
In this work, we shall unveil it first of all.
According to our study, it is clear that the logistic equation applying to penetration rates supposes the following human communication: those who do not have a prevalent stuff start to possess it shortly after they meet people already possessing it, which will be verified by numerical simulation.

Hence, a new question arises: Are we influenced by others so easily?, which is the real start-line of this paper. Therefore, we constructed a brand new model supposing more natural human communication, which implies that we extend the logistic equation.
Moreover, we adopt the total number of iPod sales as the real penetration data and, then, clarify that our model can describe the behavior much better than using the logistic equation.

\section{Penetration rate in an imitating group}

In this section, we unveil the human communication yielding the penetration rate which can be described by the logistic function. In addition, we confirm this by numerical simulation.

Let us consider a group composed of $N$ people. For this group, we shall apply the following rules: 
i) At the beginning, some people have a stuff which will diffuse in this society. ii) If those who do not possess the stuff yet ({\it non-adopters}) meet people who already possessing it ({\it adopters}) , they start to adopt it at once. iii) A non-adopter is not influenced by more than one adopter and an adopter can not influence more than one non-adopter at the same time. iv) Adopters do not part with it. 
Such a group can be realized by considering people existing on lattices whose number is $n\times n \ (<N)$. We suppose that he/she moves to one of the next lattices with the same probability at each step whose time interval is $\Delta t$. 
The third rule means that only one adopter and one non-adopter can share a lattice.
Owing to these rules, we can think of empty lattices as those who do not interact with adopters and non-adopters: people who have no interests in the stuff. Thus, we can treat a group where there are adopters, future adopters and {\em non-interested people}.

Here, we respectively set the number of adopters and non-adopters at $i$th step as $P_i$ and $Q_i$. Therefore, it is natural to consider that a probability to meet adopters or non-adopters is proportional to each number of them:
\begin{equation}
\left\{
\begin{array}{rcl}
\displaystyle \mbox{probability to meet adopters at $i$th step} &=&  \displaystyle\frac{P_i}{n^2} \\
&& \\
\displaystyle \mbox{probability to meet non-adopters at $i$th step} &=& \displaystyle\frac{Q_i}{n^2} 
\end{array}
\right. \ .
\end{equation}
Indeed, a probability to meet nobody at each step is $(n^2-P_i-Q_i)/n^2=1-N/n^2$, which also can be regarded as one to meet non-interested people.

Therefore, $({P_i}/{n^2})\times Q_i$ people of non-adopters become adopters at the next step, so that we can obtain the following recursion formulae:
\begin{align}
P_{i+1} &= P_i + \frac{P_i}{n^2}Q_i  \ ,\\
Q_{i+1} &= Q_i - \frac{P_i}{n^2}Q_i  \ .
\end{align}
We shall define the number of them at $t$ as $P(t) = P(i\cdot\Delta t) \equiv P_i$ and $Q(t) = Q(i\cdot\Delta t) \equiv Q_i$.
Here, we take the limits as $\Delta t\rightarrow 0$ and $n\rightarrow\infty$ with $n^2\Delta t$ fixed. By setting the fixed value as $a/N$ and using $P(t)+Q(t)=N$, the following differential equation is derived:
\begin{equation}
\frac{{\rm d}P(t)}{{\rm d}t} = \frac{a}{N}\left\{N-P(t)\right\}P(t) , 
\end{equation}
that is the exact logistic equation. The penetration rate $p(t)\equiv P(t)/N$ satisfies the following logistic equation
 \begin{equation}
\frac{{\rm d}p(t)}{{\rm d}t} = a\left\{1-p(t)\right\}p(t) , 
\end{equation}
and this solution can be yielded as
\begin{equation}
p(t) = \frac{1}{1+\left\{\frac{1}{p(0)}-1\right\}{\rm e}^{-at}} \ .
\end{equation}
Indeed, it is pointed out in Ref.~\cite{Mansfield61} that the imitation is essential as the human communication in a group where the change of penetration rate in time is expressed by the logistic equation.  However, the above derivation helps us to reach the deeper comprehension.

The parameter $a$, which is called the {\em coefficient of imitation} \cite{Teng02,Kumar92,Sultan90}, determines the speed of the growth of the penetration. This can be expressed as 
\begin{equation}
a = \lim_{\stackrel{\Delta t\rightarrow 0}{n\rightarrow\infty}}\frac{N}{n^2}\frac{1}{\Delta t} \ .
\end{equation}
$N/n^2$ means the population density: The larger this value is, the faster the penetration rate grows. This is reflected by the simple fact that there are many encounters in the crowded society.

Let us confirm these facts by using  a brief numerical simulation. We have $N$ players walking randomly on $n\times n$ lattices with periodic boundary condition. 
Note that this random walk is very simple and is different from that supposed on the above. In short, the rare case that a non-adopter can meet more than one adopter must occur. Furthermore, $n^2$ of this simulation  is not large. Therefore, we have done 100 times independent simulations and, then, taken the ensemble average in order to negate the contribution from such a rare case.

\begin{figure}[h]
  \begin{center}
    \begin{minipage}{14pc}
      \includegraphics[scale=0.8]{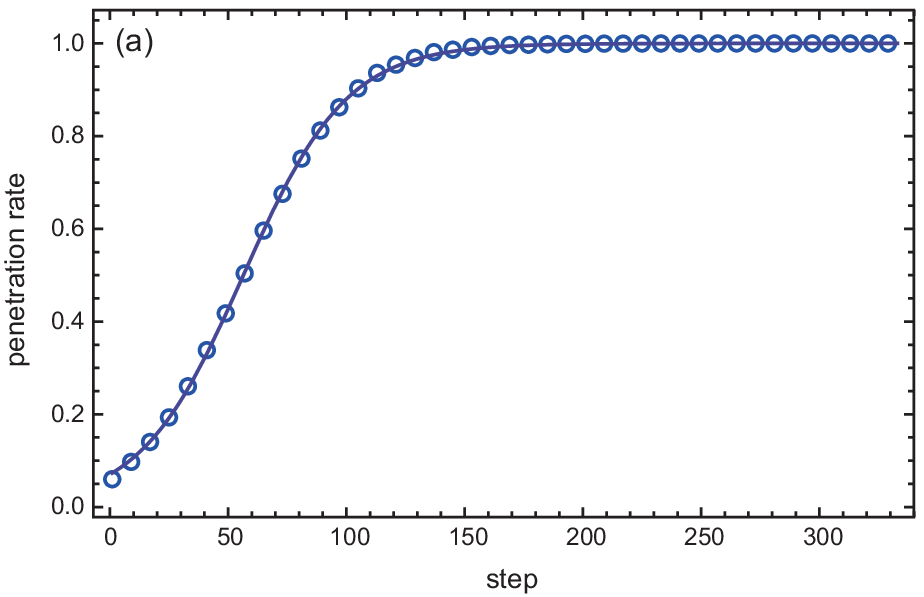}
    \end{minipage}
    \hspace{5pc}%
    \begin{minipage}{14pc}
      \includegraphics[scale=0.8]{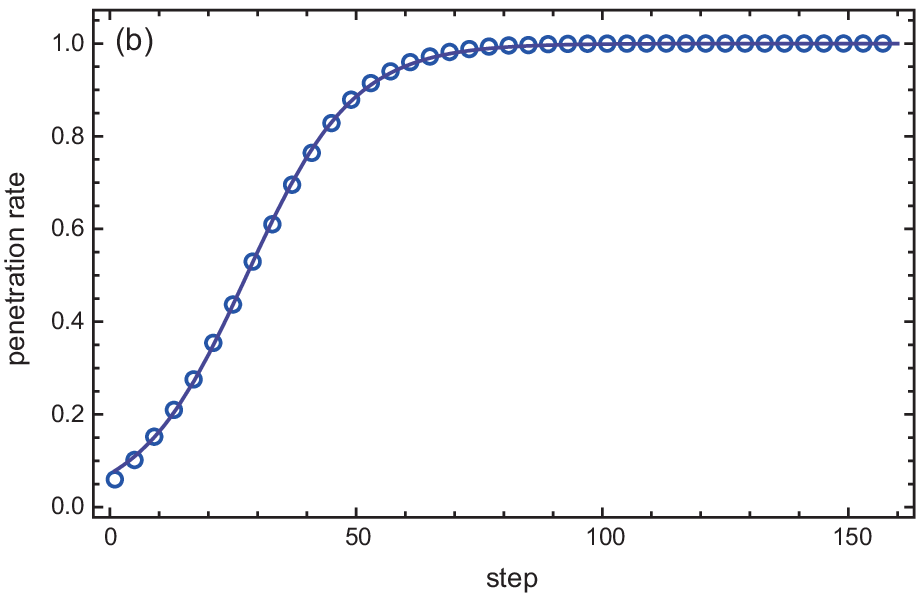}
    \end{minipage} 
    \caption{\label{Fig:simu}Penetration rates of the numerical simulations with 25 random walkers  on $n\times n$ lattices shown by circles: (a) $n=25$; (b) $n=18$. The curves expresses the logistic function $1/(1+e^{-a(t-b)})$ with an arbitrary time unit.}
  \end{center}
\end{figure}

The results with $N=25$ are show in Fig.~\ref{Fig:simu} with circles. In  Fig.~\ref{Fig:simu}(a) and (b), we set $n$ as 25 and 18, respectively. The curves mean the logistic function $1/(1+e^{-a(t-b)})$ with an arbitrary time unit. The parameters of the logistic function in Fig.~\ref{Fig:simu}(a) and (b) are $(a,b)=(0.0457,56.4)$ and $(a,b)=(0.921,27.8)$, respectively. The population density of (b) is about twice that of (a), which is consistent with the ratio of the parameter $a$ as discussed.

\section{Hierarchical logistic equation}

It is clarified in the previous section that the dynamical behavior of the penetration rate depicted by the logistic function is based on {\em imitation}: People not possessing a prevalent stuff yet go buying it shortly after they meet people already possessing it. 
Here, most of us reach the same question: But are we really like that?
In that rule, the definitive human psychology, when we imitate, is forgotten. That is, the memory, how many adopters we have met, is essential for us to start to possess the stuff

In order to integrate this feature, hereby, we shall extend the rules of the group of random walker by the following way: We set the number of people starting to process the stuff after they meet $\mu$ adopters at $i$ step  as  $Q_i^{\mu}$, in which we call $\mu$ as {\em remaining adopters number} (RAN).
Indeed, if a non-adopter, whose RAN is $\mu$, meet one of adopters, his/her RAN becomes $\mu-1$ at the next step.  
We do not alter other rules. Namely, we do not consider interactions between non-adopters despite the fact that the non-adopter gets more varied.

If the maximum of RAN is $m$, the recursion formulae exchange into 
\begin{align}
P_{i+1}          &= P_i + \frac{P_i}{n^2}Q_i^{1}  \ , \\
Q_{i+1}^{1}  &= Q_i^{1} - \frac{P_i}{n^2}Q_i^{1} + \frac{P_i}{n^2}Q_i^{2}    \ , \\
                      &\vdots  \nonumber \\
Q_{i+1}^{m} &= Q_i^{m} - \frac{P_i}{n^2}Q_i^{m} \ .
\end{align}
By the previous continuation of time and space, we can obtain the following differential equations: 
\begin{align}
\frac{{\rm d}P(t)}{{\rm d}t} &= \frac{a}{N}Q^1(t)P(t)  \ ,\\
\frac{{\rm d}Q^1(t)}{{\rm d}t} &= -  \frac{a}{N}P(t)Q^1(t) + \frac{a}{N}P(t)Q^2(t) \ , \\
                                                   &\vdots  \nonumber \\
\frac{{\rm d}Q^m(t)}{{\rm d}t} &= -  \frac{a}{N}P(t)Q^m(t)  \ , \label{eq:pq3}
\end{align}
where $Q^\mu(t) = Q^\mu(i\cdot\Delta t) \equiv Q_i^\mu$.

We shall call this the {\em hierarchical logistic equation}.
Indeed, $P(t) + Q^1(t) + \cdots + Q^{m}(t) = N$ is always conserved.

We can solve Eq.~(\ref{eq:pq3}) easily: The solution is 
\begin{equation}
  Q^m(t) = Q^m(0)\exp\left[-a\int_{0}^{t}{\rm d}t'P(t')\right] \ . 
\end{equation}
If $Q^m(0)=0$, $Q^m(t)$ is always 0. Then, the contribution of $Q^m$ into the differential equation of $Q^{m-1}$ disappears, and so $Q^{m-1}$ can be calculated similarly. Therefore, if $Q^{m}(0)=Q^{m-1}(0)=\cdots=Q^2(0)=0$, $Q^{m}(t)=Q^{m-1}(t)=\cdots=Q^2(t)=0$, which means the {\em normal} logistic equation is recovered. Namely, the hierarchical logistic equation includes the normal one.

If we use ratios of adopters and non-adopters to the total number $N$, the differential equations become
\begin{align}
\frac{{\rm d}p(t)}{{\rm d}t} &= {a}q^1(t)p(t)  \ , \\
\frac{{\rm d}q^1(t)}{{\rm d}t} &= -  {a}p(t)q^1(t) + {a}p(t)q^2(t)  \ , \\
                                                   &\vdots  \nonumber \\
\frac{{\rm d}q^m(t)}{{\rm d}t} &= -  {a}p(t)q^m(t) \ , 
\end{align}
where $q^\mu(t)\equiv Q^{\mu}(t)/N$.

\section{Fitting the iPod sales data}

\begin{figure}[h]
    \includegraphics[scale=0.65]{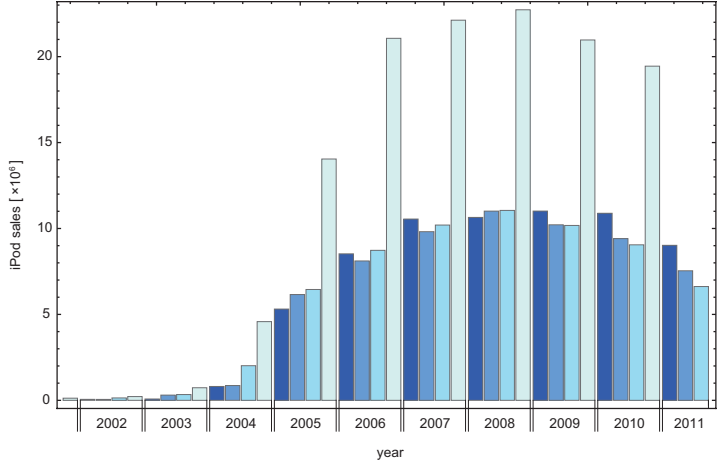}\hspace{2pc}%
    \begin{minipage}[b]{14pc}\caption{\label{Fig:iPod}(color online) iPod sales data obtained from  Apple Inc.'s official Website, http://www.apple.com/. Each year includes four quarters, February, May, August and November. The same (color) gray-level means the same quarter. The first bar is the data on November, 2001.}
  \end{minipage}
\end{figure}

Now, let us apply the hierarchical logistic equation to fitting a real data.
As this data, we shall employ the iPod, created and marketed by Apple Inc.,  sales which can be obtained from the official Website, http://www.apple.com/.
The amount of sales on a quarter is reported on the next quarter. Therefore, we consider the reported sales as on the middle of the previous quarter:  the sales reported on the first quarter as on November in the last year, the sales reported on the second quarter as on February in this year and so on\footnote{The essential is not to set the date of each data, but to set the time interval as three months}. Then, we plot the  sales as a function of time in Fig.~\ref{Fig:iPod} where the first data is in November, 2001.
As can be seen, there are six peaks after 2005. 
To our regret, the hierarchical logistic equation does not have the many peaks just like the logistic equation\footnote{However, we shall not mean that the hierarchical logistic equation has only one peak: We can discover two peaks, small and large one, with particular parameters.}.
Thus, we shall use the data from November on 2001 to May on 2006, which includes only one peak.
We treat $N$ as a fitting parameter, because the number of sales is not saturated and so we cannot obtain $N$ from the data.

\begin{figure}[h]
  \begin{center}
    \begin{minipage}{14pc}
      \includegraphics[scale=0.8]{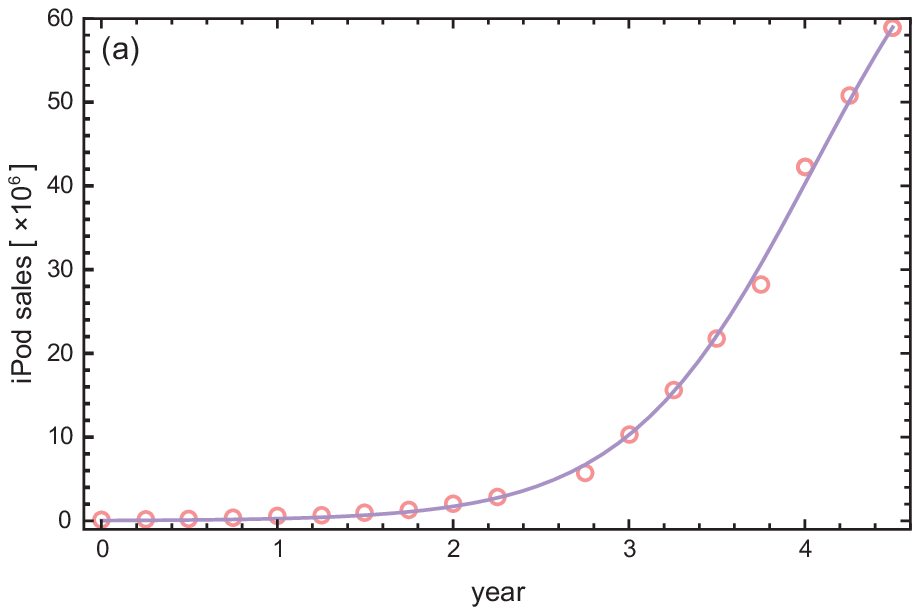}
    \end{minipage}
    \hspace{5pc}%
    \begin{minipage}{14pc}
      \includegraphics[scale=0.8]{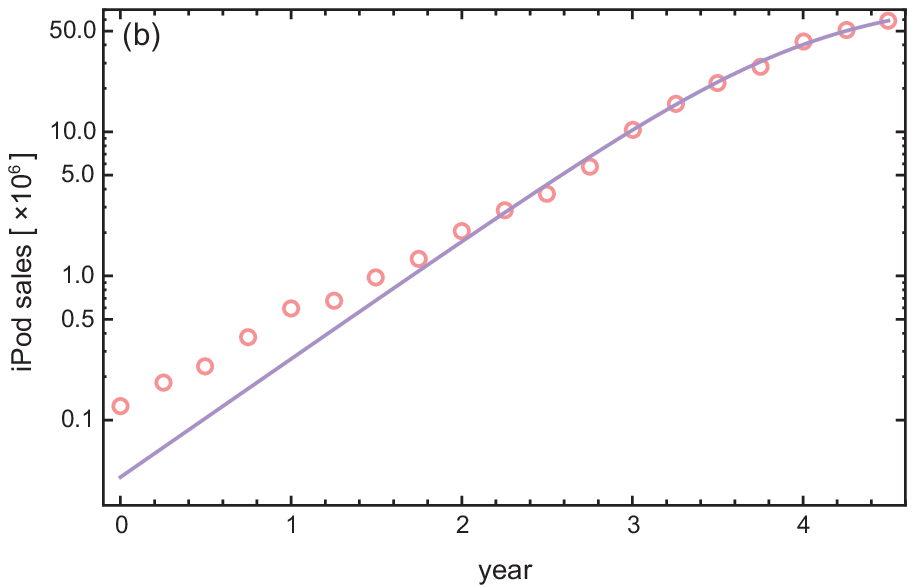}
    \end{minipage} 
    \caption{\label{Fig:m1}Cumulative iPod sales expresses by circles and the logistic function with parameters minimizing SAE.}
  \end{center}
\end{figure}

Setting November on 2001 as the origin of time, we construct the cumulative sales and, then, we fit the data with the hierarchical logistic equation.
If we minimize the residual sum of squares or the sum of the absolute value of error (SAE) when fitting this data,  the solution of the hierarchical logistic equation does not match the data with small values.
As an example, we show the cumulative iPod sales and the logistic function with parameters minimizing SAE in Fig.~\ref{Fig:m1}.
The disagreement for small values can be seen from  Fig.~\ref{Fig:m1}(b).
This results from a feature of the logistic function that $P(t)$ must be small in order to make the growth of it, $\dot{P}(t)$, small. 

\begin{figure}[h]
  \begin{center}
    \begin{minipage}{14pc}
      \includegraphics[scale=0.8]{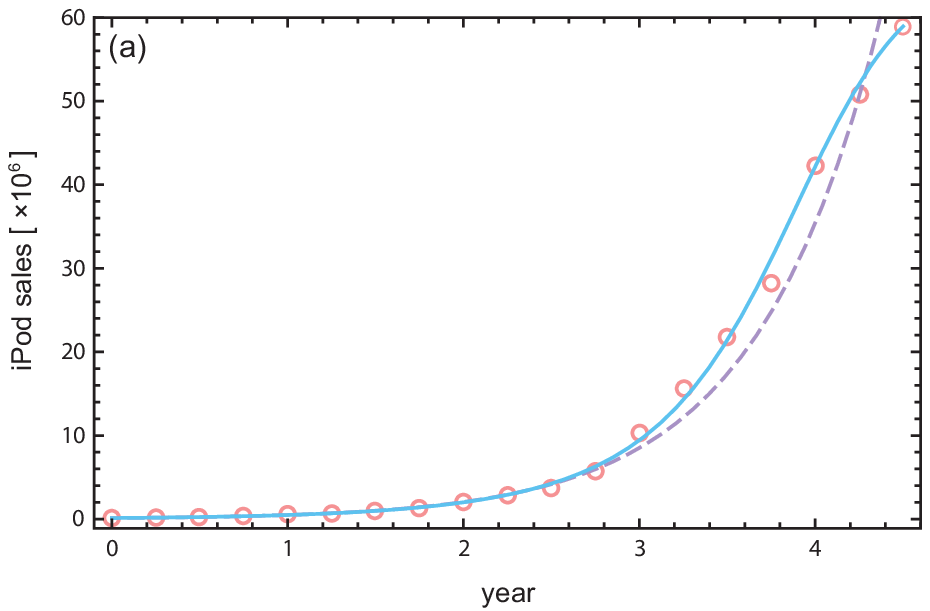}
    \end{minipage}
    \hspace{5pc}%
    \begin{minipage}{14pc}
      \includegraphics[scale=0.8]{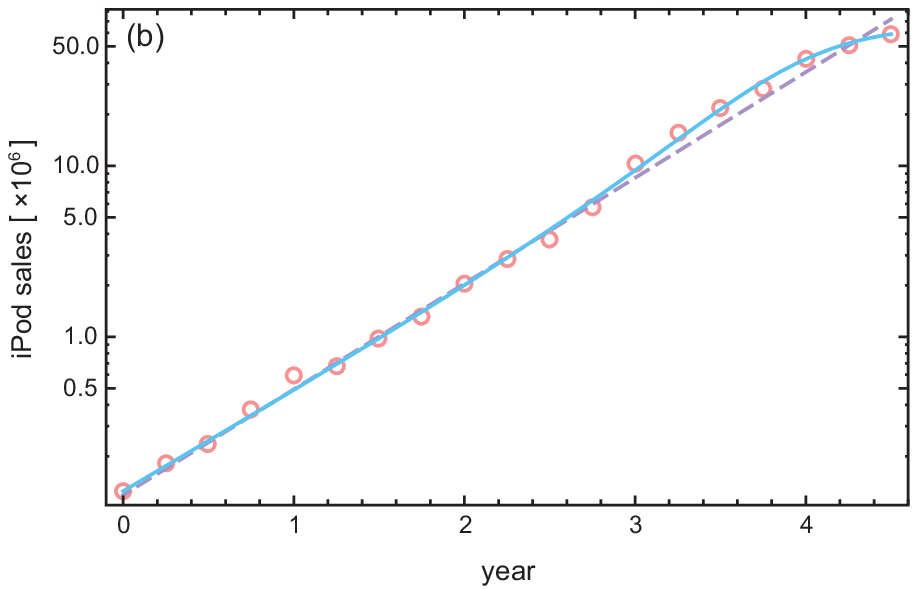}
    \end{minipage} 
    \caption{\label{Fig:m1and4}(color online) Cumulative iPod sales expresses by circles and fitting curves. The (purple) dashed and (light blue) full curves are solutions of the logistic and the hierarchical logistic equation ($m=4$) with parameters minimizing the product of SARE and SAE, respectively.}
  \end{center}
\end{figure}

Therefore, we shall minimize the product of SAE and the sum of the absolute value of relative error (SARE).
The results are show in Tab.~\ref{Tab:para}.
The parameters except in the first column minimize the product. On the contrary, those in the first column minimize SAE.
SARE diminishes with increase of $m$: SARE with $m=4$ reduces to nearly half that with $m=1$. In other words, the average relative error for the hierarchical logistic model with $m=4$, 5.3\%, is about as half as that for the logistic model, 9.7\%. 
For reference, we also show the coefficient of determination ($R^2$) in Tab.~\ref{Tab:para}, because this is employed on many papers in order to measure how well the logistic equation can approximate a real data. 
From Tab.~\ref{Tab:para}, $R^2$ is found to be not suitable for measuring the fitting accuracy: Those for the logistic and the hierarchical logistic model with $m=4$ are nearly same. However, it is very obvious that the full curves in Fig.~\ref{Fig:m1and4} expressing the solution of the hierarchical logistic equation with $m=4$ using parameters in Tab.~\ref{Tab:para} approximates the data more precisely than the curves in  Fig.~\ref{Fig:m1}.

\begin{table}
\caption{\label{Tab:para}Parameters minimizing the product of SARE and SAE (or only SAE), SAREs and $R^2$s by them.  All values are rounded to a three-digit number, and so the sum of $p(0)$ and $p^\mu(0)$ is not equal to one.}
\begin{center}
\begin{tabular*}{1.0\textwidth}{p{0.15\textwidth}p{0.21\textwidth}p{0.21\textwidth}p{0.1\textwidth}p{0.1\textwidth}p{0.1\textwidth}}
\hline
\hline
                                             & $m=1$ \ (logistic)$^*$& $m=1$ \ (logistic)      & $m=2$       & $m=3$       & $m=4$ \\
\hline
$a$                                       & $1.89$                         & $1.43$                        & $4.42$       & $4.25$       & $4.17$       \\
$N \ [\times10^6]$              & $83.7$                         & $1.28\times10^5$      & $64.6$       & $65.4$        & $66.2$      \\
$p(0)$                                   & $4.80\times10^{-4}$  & $9.26\times 10^{-7}$ & $0.00207$ & $0.00200$ & $0.00189$ \\
$q^1(0)$                               & $1.00$                         & $1.00$                         & $0.288$     & $0.311$     & $0.325$     \\
$q^2(0)$                               & $-$                              & $-$                              & $0.710$     & $0.675$     & $0.657$     \\
$q^3(0)$                               & $-$                              & $-$                              & $-$            & $0.0125$   & $0.00619$ \\
$q^4(0)$                               & $-$                              & $-$                              & $-$            & $-$             & $0.00760$  \\
\hline
SARE                                      & $4.54$                         & $1.84$                        & $1.12$      & $1.02$         & $1.01$ \\
$R^2$                                    & $0.998$                      & $0.959$                       & $0.997$   & $0.998$       & $0.998$ \\
\hline
\hline
\end{tabular*}
\end{center}
$^*$The parameters, with which the change of $P(t)(=Np(t))$ in time is shown in Fig.~\ref{Fig:m1}, minimize SAE.
\end{table}

From the parameters with $m=4$, we can find the following facts: the market size producing the first peak is about 66 million people; the ratio of the trend-conscious people,  $q^1(0)$, is about 33\%; the ratio of the cautious people, $q^2(0)$, about 66\% and the ratio of the more cautious people, $q^3(0)+q^4(0)$, is about 1\%.  

Comparing the logistic and the hierarchical logistic model, we show the result with $m=1$ and $m=4$ in Fig.~\ref{Fig:m1and4}. The (purple) dashed and the (light blue) full curves represent the logistic and the hierarchical logistic model, respectively.
The hierarchical logistic function with $m=4$ matches the data which the logistic model cannot approximate.

\section{Concluding remarks}
In this work, we have unveiled the following fact that 
the essential human communication within a group, where the dynamical behavior of the penetration rate can be approximated by the logistic function, is imitation; non-adopters start to process a prevalent stuff shortly after meeting adopters.
Indeed, this is not natural. Thereby, we  have proposed the extended logistic equation, the {\em hierarchical logistic equation}, considering the memory of the number of adopters they met.  In addition, we have applied this model to the change of iPod sales in time, and so the model has approximated the data much better than the logistic equation.
As mentioned in the previous section, the logistic equation cannot describe a slow growth as seen in iPod sales in the early 2000's, but the hierarchical logistic equation can do so. The adopters of the hierarchical logistic equation have the inner structure, resulting in the slow growth of adopters.

Wolf and Venus proposed an extended logistic equation describing a slow growth \cite{Wolf92}. In their work, they introduced the delay time, $t_i$, and multiplied  $1-\exp(-t/t_i)$ and the right hand side of the logistic equation,  Eq.~(\ref{eq:logi}). However, we emphasize that our model does not need to insert such a extra quantity.

One of our conclusions is that the essential process for a stuff to spread is imitation. It is no doubt, however, that advertisements are also essential.  The logistic model incorporating this effect is the Bass model \cite{Bass69}.
Therefore, as a future work, we shall produce the {\em hierarchical Bass model} by considering the memory on the Bass model.

\section{Acknowledgments}
We would like to thank Tomo Tanaka and members of
astrophysics laboratory at Ochanomizu University for extensive discussions.

\end{document}